\documentclass[iop,twocolumn,twocolappendix,numberedappendix]{aastex631}
\usepackage{epsfig}
\usepackage{amsmath}
\usepackage{amssymb}
\usepackage{natbib}
\usepackage{hyperref}
\usepackage{graphicx}
\usepackage{mathrsfs}
\usepackage{times}
  \usepackage[flushleft]{threeparttable}

\begin{document}
\title{PHANGS--JWST First Results: ISM structure on the turbulent Jeans scale in four disk galaxies observed by JWST and ALMA} 
\suppressAffiliations
\author[0000-0002-6118-4048]{Sharon E. Meidt}
\affiliation{Sterrenkundig Observatorium, Universiteit Gent, Krijgslaan 281 S9, B-9000 Gent, Belgium}

\author[0000-0002-5204-2259]{Erik Rosolowsky}
\affiliation{Department of Physics, University of Alberta, Edmonton, Alberta, T6G 2E1, Canada}

\author[0000-0003-0378-4667]{Jiayi~Sun}
\affiliation{Department of Physics and Astronomy, McMaster University, 1280 Main Street West, Hamilton, ON L8S 4M1, Canada}
\affiliation{Canadian Institute for Theoretical Astrophysics (CITA), University of Toronto, 60 St George Street, Toronto, ON M5S 3H8, Canada}

\author[0000-0001-9605-780X]{Eric W. Koch}
\affiliation{Center for Astrophysics | Harvard \& Smithsonian, 60 Garden St., 02138 Cambridge, MA, USA}

\author[0000-0002-0560-3172]{Ralf S.\ Klessen}
\affiliation{Universit\"{a}t Heidelberg, Zentrum f\"{u}r Astronomie, Institut f\"{u}r Theoretische Astrophysik, Albert-Ueberle-Stra{\ss}e 2, D-69120 Heidelberg, Germany}
\affiliation{Universit\"{a}t Heidelberg, Interdisziplin\"{a}res Zentrum f\"{u}r Wissenschaftliches Rechnen, Im Neuenheimer Feld 205, D-69120 Heidelberg, Germany}

\author[0000-0002-2545-1700]{Adam~K.~Leroy}
\affiliation{Department of Astronomy, The Ohio State University, 140 West 18th Avenue, Columbus, OH 43210, USA}

\author[0000-0002-3933-7677]{Eva~Schinnerer}
\affiliation{Max Planck Institute for Astronomy, K\"onigstuhl 17, 69117 Heidelberg, Germany}

\author[0000-0003-0410-4504]{Ashley.~T.~Barnes}
\affiliation{Argelander-Institut f\"{u}r Astronomie, Universit\"{a}t Bonn, Auf dem H\"{u}gel 71, 53121, Bonn, Germany}

\author[0000-0001-6708-1317]{Simon C.~O.\ Glover}
\affiliation{Universit\"{a}t Heidelberg, Zentrum f\"{u}r Astronomie, Institut f\"{u}r Theoretische Astrophysik, Albert-Ueberle-Stra{\ss}e 2, D-69120 Heidelberg, Germany}

\author[0000-0003-0946-6176]{Janice C. Lee}
\affiliation{Gemini Observatory/NSF’s NOIRLab, 950 N. Cherry Avenue, Tucson, AZ, USA}
\affiliation{Steward Observatory, University of Arizona, 933 N Cherry Ave,Tucson, AZ 85721, USA}

\author[0000-0002-5027-0135]{Arjen van der Wel}
\affiliation{Sterrenkundig Observatorium, Universiteit Gent, Krijgslaan 281 S9, B-9000 Gent, Belgium}

\author[0000-0002-7365-5791]{Elizabeth~J.~Watkins}
\affiliation{Astronomisches Rechen-Institut, Zentrum f\"{u}r Astronomie der Universit\"{a}t Heidelberg, M\"{o}nchhofstra\ss e 12-14, 69120 Heidelberg, Germany}

\author[0000-0002-0012-2142]{Thomas G. Williams}
\affiliation{Max Planck Institute for Astronomy, K\"onigstuhl 17, 69117 Heidelberg, Germany}

\author[0000-0003-0166-9745]{F. Bigiel}
\affiliation{Argelander-Institut f\"{u}r Astronomie, Universit\"{a}t Bonn, Auf dem H\"{u}gel 71, 53121, Bonn, Germany}

\author[0000-0003-0946-6176]{Médéric~Boquien}
\affiliation{Centro de Astronomía (CITEVA), Universidad de Antofagasta, Avenida Angamos 601, Antofagasta, Chile}

\author[0000-0003-4218-3944]{Guillermo A. Blanc}
\affiliation{The Observatories of the Carnegie Institution for Science, 813 Santa Barbara St., Pasadena, CA, USA}
\affiliation{Departamento de Astronom\'ia, Universidad de Chile, Camino del Observatorio 1515, Las Condes, Santiago, Chile}

\author[0000-0001-5301-1326]{Yixian Cao}
\affiliation{Max-Planck-Institut f\"ur Extraterrestrische Physik (MPE), Giessenbachstr. 1, D-85748 Garching, Germany}

\author[0000-0002-5635-5180]{M\'elanie Chevance}
\affiliation{Universit\"{a}t Heidelberg, Zentrum f\"{u}r Astronomie, Institut f\"{u}r Theoretische Astrophysik, Albert-Ueberle-Stra{\ss}e 2, D-69120 Heidelberg, Germany}
\affiliation{Cosmic Origins Of Life (COOL) Research DAO, coolresearch.io}

\author[0000-0002-5782-9093]{Daniel~A.~Dale}
\affiliation{Department of Physics and Astronomy, University of Wyoming, Laramie, WY 82071, USA}

\author[0000-0002-4755-118X]{Oleg~V.~Egorov}
\affiliation{Astronomisches Rechen-Institut, Zentrum f\"{u}r Astronomie der Universit\"{a}t Heidelberg, M\"{o}nchhofstra\ss e 12-14, 69120 Heidelberg, Germany}
\affiliation{Sternberg Astronomical Institute, Lomonosov Moscow State University, Universitetsky pr. 13, 119234 Moscow, Russia}

\author[0000-0002-6155-7166]{Eric Emsellem}
\affiliation{European Southern Observatory, Karl-Schwarzschild-Stra{\ss}e 2, 85748 Garching, Germany}
\affiliation{Univ Lyon, Univ Lyon1, ENS de Lyon, CNRS, Centre de Recherche Astrophysique de Lyon UMR5574, F-69230 Saint-Genis-Laval France}

\author[0000-0002-3247-5321]{Kathryn~Grasha}
\affiliation{Research School of Astronomy and Astrophysics, Australian National University, Canberra, ACT 2611, Australia}   
\affiliation{ARC Centre of Excellence for All Sky Astrophysics in 3 Dimensions (ASTRO 3D), Australia}

\author[0000-0001-9656-7682]{Jonathan~D.~Henshaw}
\affiliation{Astrophysics Research Institute, Liverpool John Moores University, 146 Brownlow Hill, Liverpool L3 5RF, UK}
\affiliation{Max Planck Institute for Astronomy, K\"onigstuhl 17, 69117 Heidelberg, Germany}

\author[0000-0002-8804-0212]{J.~M.~Diederik~Kruijssen}
\affiliation{Cosmic Origins Of Life (COOL) Research DAO, coolresearch.io}

\author[0000-0003-3917-6460]{Kirsten L. Larson}
\affiliation{AURA for the European Space Agency (ESA), Space Telescope Science Institute, 3700 San Martin Drive, Baltimore, MD 21218, USA}

\author[0000-0001-9773-7479]{Daizhong Liu}
\affiliation{Max-Planck-Institut f\"ur Extraterrestrische Physik (MPE), Giessenbachstr. 1, D-85748 Garching, Germany}

\author[0000-0001-7089-7325]{Eric J.\,Murphy}
\affiliation{National Radio Astronomy Observatory, 520 Edgemont Road, Charlottesville, VA 22903, USA}

\author[0000-0003-3061-6546]{Jérôme Pety}
\affiliation{IRAM, 300 rue de la Piscine, 38400 Saint Martin d'H\`eres, France}
\affiliation{LERMA, Observatoire de Paris, PSL Research University, CNRS, Sorbonne Universit\'es, 75014 Paris}

\author[0000-0002-0472-1011]{Miguel~Querejeta}
\affiliation{Observatorio Astron\'{o}mico Nacional (IGN), C/Alfonso XII, 3, E-28014 Madrid, Spain}

\author[0000-0002-2501-9328]{Toshiki Saito}
\affiliation{National Astronomical Observatory of Japan, 2-21-1 Osawa, Mitaka, Tokyo, 181-8588, Japan}

\author[0000-0002-4378-8534]{Karin M. Sandstrom}
\affiliation{Center for Astrophysics \& Space Sciences, University of California, San Diego, 9500 Gilman Drive, San Diego, CA 92093, USA}

\author[0000-0002-0820-1814]{Rowan J. Smith}
\affiliation{Jodrell Bank Centre for Astrophysics, Department of Physics and Astronomy, University of Manchester, Oxford Road, Manchester M13 9PL, UK}

\author[0000-0001-6113-6241]{Mattia C. Sormani}
\affiliation{Universit\"{a}t Heidelberg, Zentrum f\"{u}r Astronomie, Institut f\"{u}r Theoretische Astrophysik, Albert-Ueberle-Stra{\ss}e 2, D-69120 Heidelberg, Germany}

\author[0000-0002-8528-7340]{David A. Thilker}
\affiliation{Department of Physics and Astronomy, The Johns Hopkins University, Baltimore, MD 21218, USA}

\begin{abstract}
JWST/MIRI imaging of the nearby galaxies IC 5332, NGC 628, NGC 1365 and NGC 7496 from PHANGS reveals a richness of gas structures that in each case form a quasi-regular network of interconnected filaments, shells and voids.  We examine whether this multi-scale network of structure is consistent with the fragmentation of the gas disk through gravitational instability.  We use FilFinder to detect the web of filamentary features in each galaxy and determine their characteristic radial and azimuthal spacings.  These spacings are then compared to estimates of the most Toomre-unstable length (a few kpc), the turbulent Jeans length (a few hundred pc) and the disk scale height (tens of pc) reconstructed using PHANGS-ALMA observations of the molecular gas as a dynamical tracer.  Our analysis of the four galaxies targeted in this work indicates that Jeans-scale structure is pervasive.  Future work will be essential for determining how the structure observed in gas disks impacts not only the rate and location of star formation but also how stellar feedback interacts positively or negatively with the surrounding multi-phase gas reservoir.  
\end{abstract}
\section{Introduction\label{sec:intro}}
\setcounter{footnote}{0}
JWST MIRI imaging of nearby galaxies offers an unprecedented high-resolution, high-sensitivity view of the structure and organization of the interstellar medium 
out of which new stars form.  The emission from PAHs captured in the 7.7 and 11.3 micron bands, in particular, is used not only as a tracer of star formation \citep[e.g.][]{kenn09, calzetti10} but also of molecular gas (e.g. \citealt{regan06,crocker13,leroy21b,chown21}; and see \citealt{LEROY1_PHANGSJWST,SANDSTROM1_PHANGSJWST,LEE_PHANGSJWST}).   Indeed, these small complex molecules probe the UV radiation sources responsible for their excitation \citep{Draine21}. Away from these sources, PAHs act as exceptional tracers of the cold dense gas phase thanks to a strong dynamical coupling through drag forces \citep{draine04}.  

\begin{figure*}[t]
\begin{center}
\begin{tabular}{cc}
\includegraphics[width=.45\linewidth]{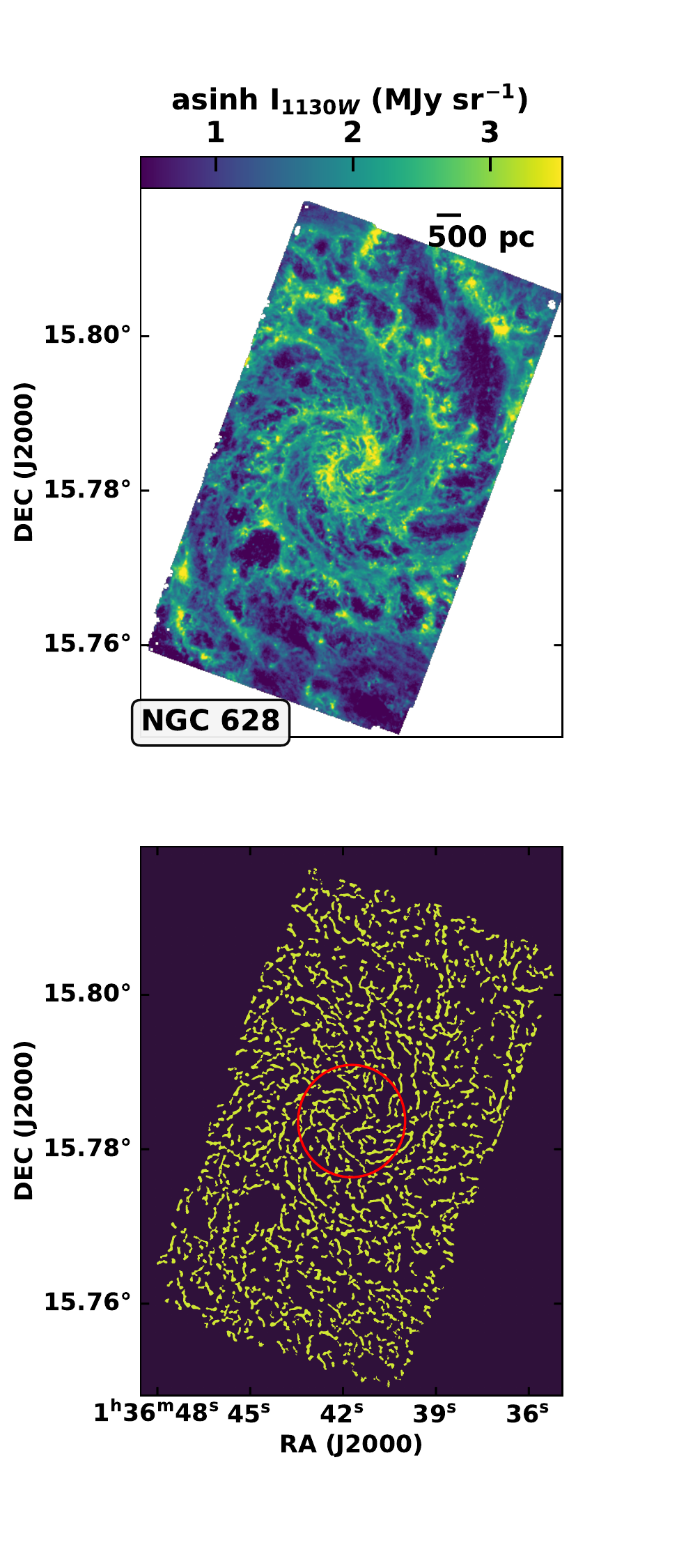}&
\hspace*{-0.25cm}\includegraphics[width=.475\linewidth]{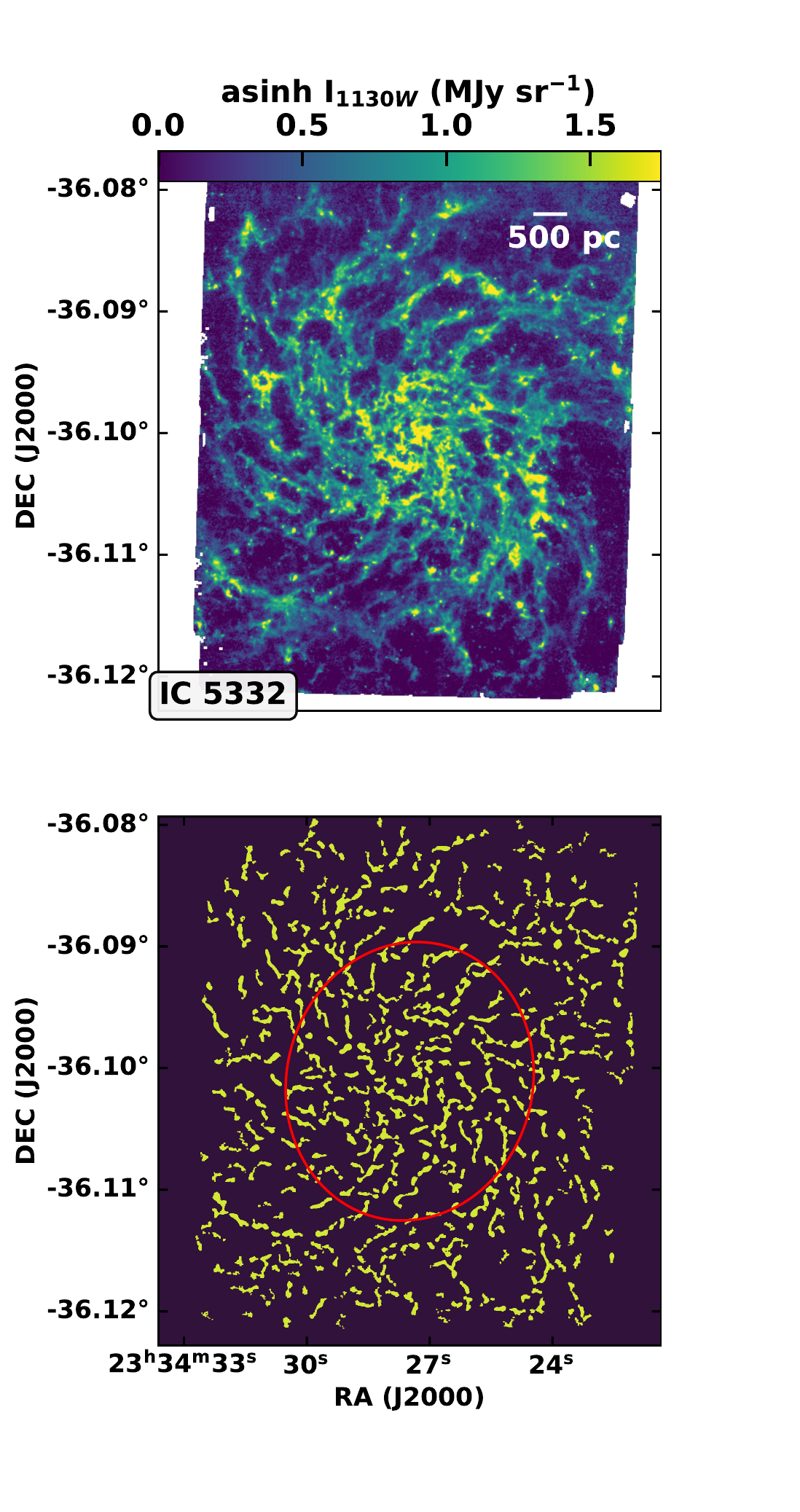}

\end{tabular}
\end{center}
\caption{(Top) JWST/MIRI 11.3 micron images and (bottom) FilFinder masks (see $\S$ \ref{sec:FilFinder}) for two PHANGS-JWST targets, NGC~0628 (left) and IC~5332 (right).  The MIRI images are displayed on a asinh stretch.  The physical scale is indicated by a white bar in the left panels, while the red ellipse in the right panels positioned at a galactocentric radius $R$=2~kpc marks the orientation of elliptical annuli in the azimuthal spacings of the filaments identified in these two galaxies.  \\
 }
\label{fig:maps1}
\end{figure*}

\begin{figure*}[t]
\begin{center}
\begin{tabular}{cc}
\hspace*{-1.5cm}\includegraphics[width=.75\linewidth]{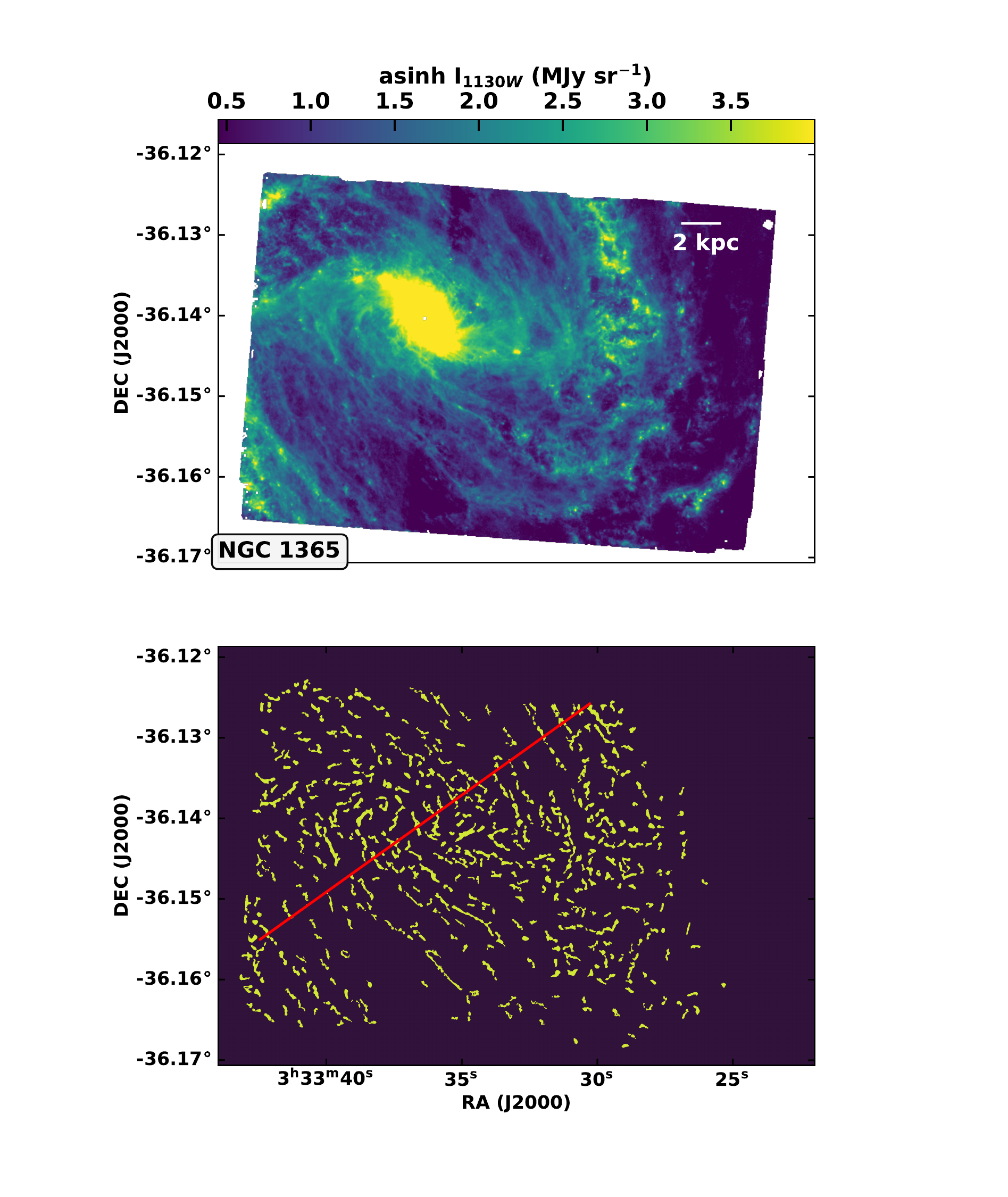}&
\hspace*{-1.8cm}\includegraphics[width=.45\linewidth]{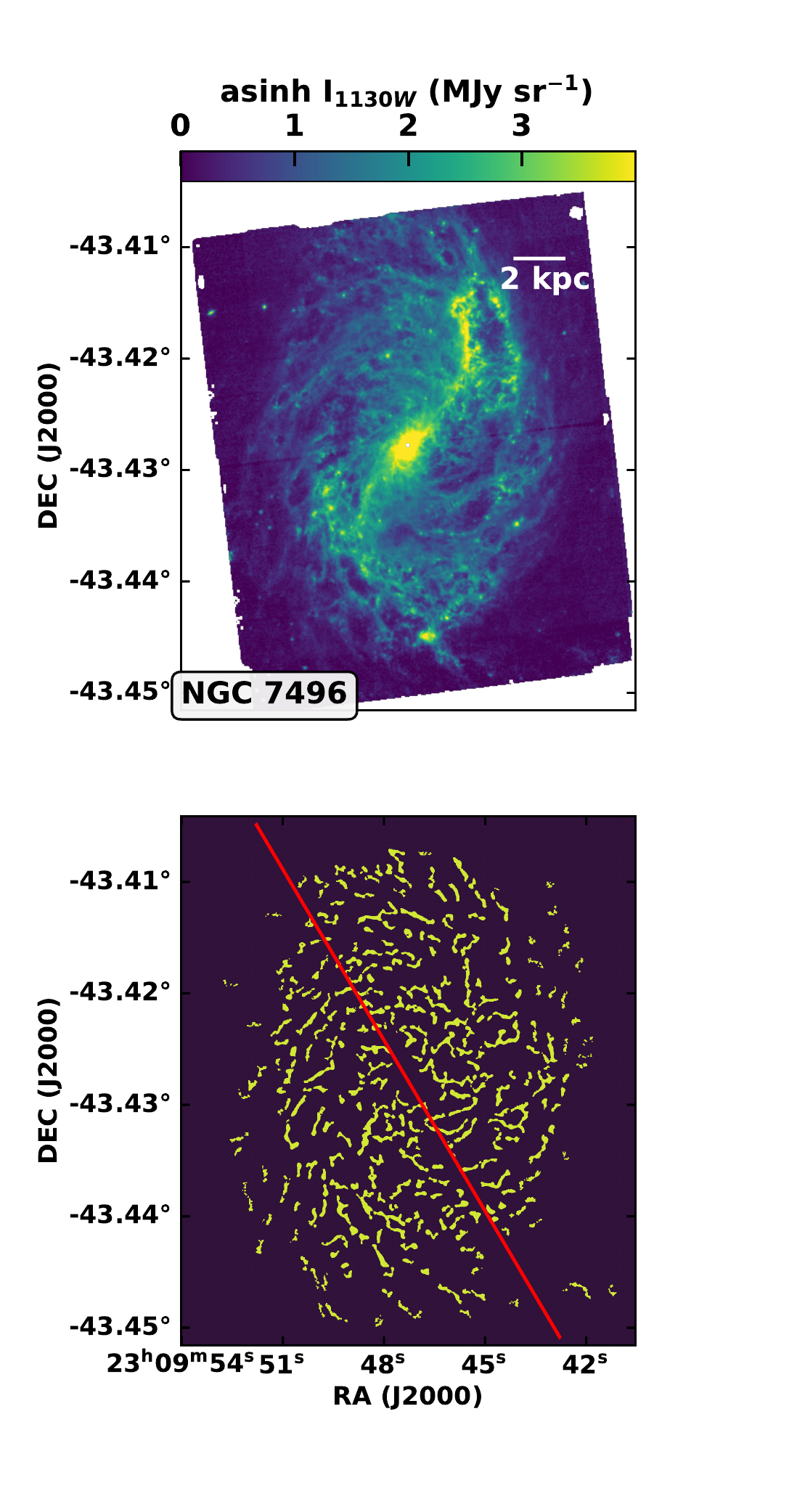}
\end{tabular}
\end{center}
\caption{(Top) JWST/MIRI 11.3 micron images and (bottom) FilFinder masks (see $\S$ \ref{sec:FilFinder}) for two PHANGS-JWST targets, NGC~1365 (left) and NGC~7496 (right).  The MIRI images are displayed on a asinh stretch. The physical scale is indicated by a white bar in the left panels, while the red lines in the right panels illustrate the orientation of one of the segments along which radial spacings are measured in these two galaxies. (deprojected).\\
 }
\label{fig:maps2}
\end{figure*}

One of the striking aspects of the new images is that they sharpen and extend our view of gas morphology traced by CO \citep[see also][]{LEROY1_PHANGSJWST}, revealing the presence of an unambiguous network of highly structured features -- consisting of filaments, shells and voids -- across multiple scales (see Figures \ref{fig:maps1} and \ref{fig:maps2}).  In some regions, this web of features appears built from individual or overlapping shells and bubbles where recent star formation has had a visible influence on the local gas morphology \citep{WATKINS_PHANGSJWST,BARNES_PHANGSJWST}, and as shown by \cite{WATKINS_PHANGSJWST}, the JWST/MIRI imaging exhibit bubbles ranging in size from tens to hundreds of pc.  As we will show here \citep[see also][]{THILKER_PHANGSJWST}, connecting and underlying these bubbles, there is a more extended and  highly regular, quasi-periodic pattern of filamentary features that exhibit similar lengths, widths, spacings and orientations.  These filaments range in size from $\sim$ 50 pc to 1 kpc or longer.  In the case of NGC~628, these narrow features are so extended that they traverse the distance between the two prominent spiral arms more clearly than has been observed with any other high resolution ISM tracer.

In the Milky Way (MW), filamentary features have been observed over an enormous range of spatial scales, from several hundreds of pc all the way down to 0.01 pc, and there are a number of theories for their formation on various scales \citep[see review by][]{hacar22}.  The set of `giant filaments' often linked to the MW's spiral arms and mid-plane (e.g. \citealt{ragan14, abreu16, zucker15, wang15, li16, veena21}; see also \citealt{colombo21,soler22}) are most closely analogous to the extragalactic filaments observed with JWST.  The filaments in Figures \ref{fig:maps1} and \ref{fig:maps2}, however, are markedly ubiquitous, present within and between spiral arms, at small and large galactocentric radii, and in galaxies with diverse star formation rates, masses, gas contents and morphologies \citep[see Table \ref{tab:props} and][]{LEE_PHANGSJWST}. 

A possible clue to the origin of the multi-scale web structure observed in extragalactic targets is that the interarm features in NGC~628, which trace back towards the arms, match identically onto the regular spur and feather features observed in CO emission or dust extinction in the vicinity of the arms (\citealt{lavigne}; see also \citealt{THILKER_PHANGSJWST} and \citealt{WILLIAMS_PHANGSJWST}).  The correspondence with the feathers cataloged by \cite{lavigne} is striking.  As examined in that study, the emergence of such gas structures is thought to be highly sensitive to the magneto-hydrodynamics of gas flow in a spiral potential (e.g. \citealt{ko06,WadaKoda,dobbsbonnell}; and see \citealt{Mandowara}).

At a more basic level, the formation of filamentary features also seems tied to the fundamental process of disk fragmentation mediated by the interplay between gas self-gravity, galactic rotation and turbulent gas pressure.  This might also be expected to generate a quasi-periodic/semi-regular pattern (\citealt{toomre, GLB, meidt22}; see also \citealt{henshaw20} and \citealt{utreras}) similar to what is observed.  An example is given by the MIRI imaging of the molecule-poor galaxy IC 5332, which lacks a strong underlying large-scale pattern and is only weakly forming massive stars, but nonetheless exhibits a filamentary network that bears a striking resemblance to the web in NGC~628.  It is also morphologically similar to the cold gas in the simulation of multi-phase disk fragmentation studied by \cite{WadaNorman}.  More recent simulations have demonstrated the important role of differential rotation, which stretches and elongates overdensities, in the formation of long filaments \citep{smith14, duarte17, smith20}.

The MIRI observations of the barred galaxies NGC 1365 and NGC 7496 offer further insight into the formation/nature of filamentary web structures. In both cases, elongated ring-like features several kpc in diameter centered on each galaxy's center are observed, although the structures in the bar in NGC 1365 are far more spatially coherent and regularly spaced than in NGC 7496.  Considering that the latter's bar contains more plentiful star formation activity, the remarkable  coherence of NGC 1365's elongated features seems to owe as much to the absence of strong star formation in the bar region as to the strength of motion in the galactic potential.  

In this paper we measure the spacing of features in the dusty webs of NGC~628, IC~5332, NGC~1365 and NGC~7496 imaged by PHANGS with JWST/MIRI and examine whether they are consistent with empirical estimates of the Toomre length, the turbulent Jeans length and the disk scale height constructed using PHANGS-ALMA CO(2-1) observations of each galaxy \citep{leroy21a}.  The first two lengths represent predictions for the sizes of fragments formed through gravitational instability in rotating disks \citep{toomre, GLB, elm87,ko06, meidt22}, while the disk scale height is a naturally imposed length scale for the
system that delimits a boundary for the expansion of feedback-driven bubbles \citep[e.g.][]{maclow88,kim17,fielding18,orr22}. 
The basic approach adopted in this work can serve as a basis for future efforts to study the properties of extragalactic filamentary structures and quantify the interplay between star formation feedback and galaxy dynamics in shaping the cold gas reservoir over time.  

\begin{table*}[t]
\begin{center}
\caption{Properties of the four targeted galaxies }\label{tab:props}
\begin{threeparttable}
\begin{tabular}{rcccc}
\hline\hline
 &IC 5332 & NGC 628 & NGC 1365 & NGC 7496\\
\hline
Distance (Mpc)&9.0&9.8&19.6&18.7\\
i (deg) &27&9&55&36\\
PA (deg)&74&21&201&194\\
log Stellar mass (M$_\odot$)&9.67&10.34&10.99&10.0\\
log SFR (M$_\odot$ yr$^{-1}$)&-0.39&0.24&1.24&0.35\\
\hline
$\Sigma_{mol}$(M$_\odot$ pc$^{-2}$)\tnote{a}&0.5 (0.3 -- 1.4)&11.2 (5.8 -- 19.3)&8 (3.0 -- 59.0) &7.4 (2.0 -- 17.8)\\
$\sigma_{mol}$(km s$^{-1}$)\tnote{a}&1.25\tnote{d} &4.7 (3.9 -- 5.5)&8.7 (3.5 -- 23)&4.3 (3.4 -- 6.2) \\
$\Sigma_{mol}/\Sigma_{atomic}$\tnote{b}& & & &\\
\hline
$\lambda_T$(pc)\tnote{a}&1523 (594 -- 2678) &9638 (4584 -- 14367)&5397 (625 -- 8107)&37782 (4504 -- 39725) \\
$\lambda_J$(pc)\tnote{c}&673 (404 -- 906) &590 (445 -- 708)&1145 (712 -- 1603) &742 (529 -- 966) \\
$\lambda_{fil}$(pc)\tnote{d}&660 (420 -- 888) &388 (335 -- 437) &1051 (647 -- 1285) &547 (434 -- 652) \\

\hline
 
\end{tabular}
\tablecomments{Distances from the compilation of \cite{anand21}.  Stellar masses and star formation rates determined by \cite{leroy21a}.  Position angles and inclination angles from \cite{lang20} where available and \cite{leroy21a} for IC 5332.}
 \begin{tablenotes}
\item[a]The mean of the values measured in all hexagonal apertures (see $\S$ \ref{sec:almadata}) sampling the analysis region.  Indicated ranges denote the 16th and 84th percentiles of the measurements. 
\item[c]The mean of the values measured in all hexagonal apertures (see $\S$ \ref{sec:almadata}) assuming a fixed gas scale height $h$=100 pc.  Indicated ranges denote the 16th and 84th percentiles of the measurements. 
\item[d]The mean of all radially binned azimuthal spacings measured for IC 5332 and NGC 628 or the mean of all azimuthally binned radial spacings in NGC 1365 and NGC 7496, as plotted in Figure \ref{fig:spacings}. Indicated ranges denote the 16th and 84th percentiles of the measurements. 
\item[d]A single global value is chosen for the analysis (see text).
 \end{tablenotes}
\end{threeparttable}
\end{center}
\end{table*}


\section{The data}
\subsection{PHANGS-JWST imaging}\label{sec:jwstdata}
In this paper we study the first four targets observed to date as part of the PHANGS-JWST cycle 1 treasury project (ID 02107): IC 5332, NGC 628, NGC 1365 and NGC 7496 \citep{LEE_PHANGSJWST}. IC 5332 is a flocculent dwarf spiral, NGC 628 is a moderate-mass grand design spiral, NGC 1365 is a massive, strongly-barred spiral galaxy, and NGC 7496 is a moderate-mass barred spiral \citep{LEE_PHANGSJWST}. 

Our focus for this pilot study is on MIRI F1130W imaging (0.36'' full-width-half-maximum resolution) where the sensitivity to low column densities is highest \citep{LEROY2_PHANGSJWST}.  Images show exceptional sensitivity to spatially extended filamentary features of interest in this work (see Figures \ref{fig:maps1} and \ref{fig:maps2}).  This band also shows a tight empirical correlation with the CO emission in these four targets that may indicate that the PAH emission in this band is a sensitive tracer of multiple gas phases  \citep{LEROY1_PHANGSJWST}.  These data were processed together with the suite of NIRCam (F200W, F300M, F335M and F360M) and MIRI (F770W, F1000W, F1130W and F2100W) imaging obtained for each target \citep[see][]{LEE_PHANGSJWST}.  The field-of-view in each case overlaps existing coverage from HST \citep{lee22}, VLT-MUSE \citep{emsellem22} and ALMA \citep{leroy21b} assembled by the PHANGS survey. 

Table \ref{tab:props} lists a number of relevant properties for the four target galaxies \citep[see also][]{LEE_PHANGSJWST}.  We note here that, at the distances of our targets, 0.36'' corresponds to 16-34~pc.

\subsection{PHANGS-ALMA}\label{sec:almadata}
We use PHANGS-ALMA CO (2-1) \citep{leroy21a,leroy21b} as our primary tracer of the dynamics of the gas, its global rotation, turbulent velocity dispersion and mass surface density.  These are necessary inputs for estimating the Jeans length and the Toomre length (see $\S$ \ref{sec:jeanstoomre}) throughout the gas disk.  Although it may be a promising area in the future, CO is not employed in this work as a tracer of the properties of individual dusty filaments.

In three of the four targets, the gas disk covered by the MIRI imaging is dominated by the molecular gas traced by CO.  In NGC~628, the azimuthally-averaged molecular gas mass fraction is $>$70\% over the MIRI field of view using the atomic gas traced by resolved HI observations available for NGC~628 \citep[see][]{sun22}.   A similar fraction of the inner ISM is molecular in NGC 7496 and this raises to almost 90\% across the targeted field of view in NGC~1365, adopting a representative HI surface density of 10 $M_\odot$ pc$^{-2}$ \citep{bigielblitz12}.  For these three targets, we can thus directly leverage the gas properties measured by PHANGS-ALMA.  In the case of the molecule-poor galaxy IC 5332, we estimate that the ISM is roughly 10\% molecular and thus we supplement CO constraints with additional information, as described more below. 

For NGC~628, NGC~1365, and NGC~7496, we adopt the rotation curve measured from the CO velocity field by \cite{lang20} as our estimate of the circular velocity $V_c$. For IC~5332, we use the semi-empirical rotation curve model implied by this galaxy's stellar mass \citep[see][]{leroy21b}, following \cite{meidt18}. Each rotation curve is interpolated onto a finer radial grid so that each pixel (and resolution element) in the JWST map has an estimate of the circular velocity $V_c$ associated with it.  This approach may lead to mis-estimation of $V_c$ where strong non-circular motions are present and contribute to the measured rotation curve (within the barred regions of NGC~1365 and NGC~7496, for example).  The result is that the Toomre length may be underestimated at these radii (see $\S$ \ref{sec:jeanstoomre}).  

For the gas properties (velocity dispersion and surface density), where possible we adopt the values tabulated for the molecular gas in the PHANGS mega-tables recently presented by \cite{sun22}.  Our interest is in approximating the Jeans and Toomre lengths in the gas initially, prior to fragmentation.  We therefore adopt the `large-scale' area-weighted average molecular gas surface density measured in 1-kpc hexagonal apertures for NGC 628, NGC 1365 and NGC 7496.  In a given hex aperture, these values are calculated from the total flux traced in the 12m+7m+tp CO(2-1) 150~pc-scale integrated intensity (moment-0) map within the hex boundary, divided by the area of the aperture.  \citep[See][for the adopted conversion factor $\alpha_{\mathrm CO}$ and other details.]{sun22} For IC 5332, the flux recovery is considerably higher in the 7m+tp CO(2-1) cube (compared to the 12m+7m+tp cube), and so we construct our `large-scale' area-weighted molecular surface densities from the 7m+tp moment-0 map convolved to 1.5 kpc resolution, sampling the convolved map at the center of each hex aperture.  We then apply an $\alpha_{\mathrm CO}$ appropriate for each aperture following the prescription of \cite{sun22}.

As a best approximation for the initial velocity dispersion in the cold gas, where possible we select the average cloud-scale velocity dispersion in each hexagonal aperture (as opposed to a larger-scale measure, which would likely reflect unresolved galactic motion).  This assumes that principal fragmentation happens at/near the cloud scale rather than above it; any scale-dependent evolution in turbulent velocity dispersion is assumed to occur below the cloud scale, e.g. where it can lead to secondary fragmentation and eventually lead to star formation.  An alternative choice would be to adopt a single value representative of the neutral gas as a whole, e.g. $\sigma_{gas}\approx$ 11 km s$^{-1}$ following  \cite{leroy08} and others. We comment on the impact of this choice later in $\S$ \ref{sec:results}.  

As our fiducial estimate of the molecular gas velocity dispersion in each aperture, we choose the average velocity dispersion measured specifically on 150~pc scales from the 12m+7m+tp CO(2-1) cubes for NGC 628, NGC 1365 and NGC 7496  \citep[see][for details]{sun22}. This allows us to consider the same physical scale in all galaxies at the highest resolution possible (given the distance of the most distant target studied here).  For IC 5332, we expect the velocity dispersion measured from the 7m+tp moment-2 map at 1.5 kpc resolution to contain unresolved bulk motion.  We therefore select the minimum velocity dispersion $\sigma=1.25$ km s$^{-1}$ as a realistic upper bound on the velocity dispersion in the molecular gas at all  locations in the map of IC~5332.

In all four targets, we use these measurements of the molecular gas surface density and velocity dispersion as our estimates of total gas surface density and velocity dispersion, but also explore the addition of an atomic gas component in the case of the molecule-poor galaxy IC~5332 \citep{LEROY1_PHANGSJWST}, where it is most relevant.  In this case, the total gas surface density is calculated by combining the molecular surface density with an atomic gas surface density that is assumed to exponentially decline from a maximum value of 10 M$_{\odot}$ pc$^{-2}$ with a scale length of 3.6 kpc.  The integrated mass of atomic gas then equals value estimated for this galaxy \citep{leroy21b}.  When the atomic gas component included, the gas velocity dispersion is assigned a value of 7 km s$^{-1}$ \citep[e.g.][]{leroy08}. 

Supplementing this set of measured gas properties, in each hexagonal 1-kpc mega-table aperture we also use the tabulated stellar mass surface density and the stellar scale height $z_0$ assigned uniquely to each galaxy according to an empirical scaling relation between $z_0$ and disk scale length \citep[see][and references therein]{sun20}.

\section{The Measurements}
\subsection{The Characteristic Filament Separation Length}
\subsubsection{Identifying Filaments}\label{sec:FilFinder}
As our goal is to examine consistency with structure formation through gravitational instability, to analyze the JWST images we use an approach most sensitive to the filamentary features deemed most likely to be consistent with fragmentation.  Filaments are identified using FilFinder \citep{KochRosolowsky}.  For this pilot study, we have chosen to apply FilFinder as uniformly as possible, with minimal tuning of the input parameters from galaxy to galaxy.  This is to emphasize the similarity of the main filamentary features both throughout a given galaxy and between targets.  Future studies may find it useful to optimize FilFinder's input parameters on a galaxy-by-galaxy or region-by-region basis, for in-depth characterizations of  filaments, small bubbles and other shell-like features.

The input parameters are initially tuned using the prominent filament features in NGC~628 as the prototype.  Extracted features are selected to have widths not below 50 pc (roughly 3 resolution elements at the distance of NGC 628) and lengths not below 100 pc, although in the two most distant barred galaxies NGC 1365 and NGC 7496, this minimum length is increased to 500 pc, selected to avoid mildly resolved clumps not the focus of this work.  With these parameters, many visibly long filaments in all four of the targets are segmented by FilFinder, but the identified structures are suitable for our goal of measuring the spacing of structures rather than characterizing their individual lengths or geometries.

The number and basic footprints of identified structures are found to be surprisingly robust to modest changes in the primary FilFinder parameters, with some small changes in filament edges possible.  Indeed, applying FilFinder on multiple scales to the full suite of MIRI images for NGC 628, \cite{THILKER_PHANGSJWST} find that the filaments in FilFinder masks are largely stable below 100 pc scales, implying that changes in FilFinder parameters lead to identification of mostly comparable structures on these scales.  

FilFinder is slightly more sensitive to the choice of intensity threshold above which filaments are selected, which typically depends on the structures of greatest interest in a given map, such as faint interarm filaments or brighter spiral features.  To lend objectivity to this choice, in this work FilFinder is applied to `unsharp mask' versions of the original images, created as the difference between the native image and a smoothed version calculated via Gaussian convolution with a sigma of 1.1" (10 image pixels or roughly 3 resolution elements; i.e. adopting a fixed angular rather than physical smoothing scale).  This strategy was found to be an effective way to remove the need for varying thresholds within and between galaxies (due to underlying (radial) intensity gradients) while also enhancing the contrast of low brightness interarm filaments often found in interarm regions. The enhanced appearance of filaments in these images also makes it easier to inspect the physical structures identified in the FilFinder masks.  (With selective tuning of the intensity threshold, however, FilFinder returns similar filaments when applied to the native images.)

For these `unsharp mask' images, the global threshold for filament identification was chosen to retrieve the visually prominent filamentary features and avoid other elongated chain-like structures and knots sometimes present within larger-scale voids, i.e. tracing obscured star formation. The `unsharp mask' threshold level is set to 0.03 MJy/sr  in NGC~628, IC~5332 and NGC~7496.  In NGC~1365, the threshold is raised to 0.2 MJy/sr to avoid selecting substructures in the principle filamentary features of interest. 

It should be noted that unsharp masking reduces the appearance of structure below the chosen smoothing scale $\theta$. However, our testing suggests that this is less important for the identified filaments than the chosen FilFinder parameters.  Variations in the unsharp mask smoothing scale tend to prompt changes in two primary features: the widths of structures remaining in the unsharp mask image and the brightness level of the features. For example, reducing the smoothing scale from $\theta_1$ to $\theta_2$ results in structures that are narrower versions of the structures in the $\theta_1$ map that are also slightly lower in brightness.  The $\theta_2$ maps can also show an increased number of clumps and short segments detached from the main web (i.e. which were below the $\theta_1$ smoothing scale).  For the purposes of selecting the filaments that build the main web-network, however, FilFinder's adaptive threshold scale and axis ratio parameters can be chosen to avoid clumps in the $\theta_2$ maps. Provided the adaptive threshold scale and brightness threshold parameters are also adjusted to match the new properties of the narrower filaments in the $\theta_2$ map, we have found that FilFinder returns a very similar set of structures as in the nominal $\theta_1$ case.
This is consistent with previous results where FilFinder recovers high signal-to-noise structure for reasonable ranges of input parameters \citep{KochRosolowsky,green17}.

Figures \ref{fig:maps1} and \ref{fig:maps2} show the FilFinder filament mask calculated from the MIRI image for each target. Without considerable tuning, FilFinder returns masks that more than adequately capture the filaments of interest, on tens to hundreds of pc scales, and avoid knots of obscured star formation in voids and along the filaments themselves.  It should be noted that there are filaments or other elongated structures that lie at a lower intensity level than the spatially extended web-like structure that is the focus of this work. Those objects, which tend to be shorter or narrower than our selected FilFinder filaments, do not contribute to the visually striking filamentary web and are assumed to originate with a different mechanism.  We avoid those objects to provide a clean test for the formation of the main web.  

\begin{figure*}[t]
\begin{center}
\begin{tabular}{cc}
\hspace*{-.5cm}\includegraphics[width=.5\linewidth]{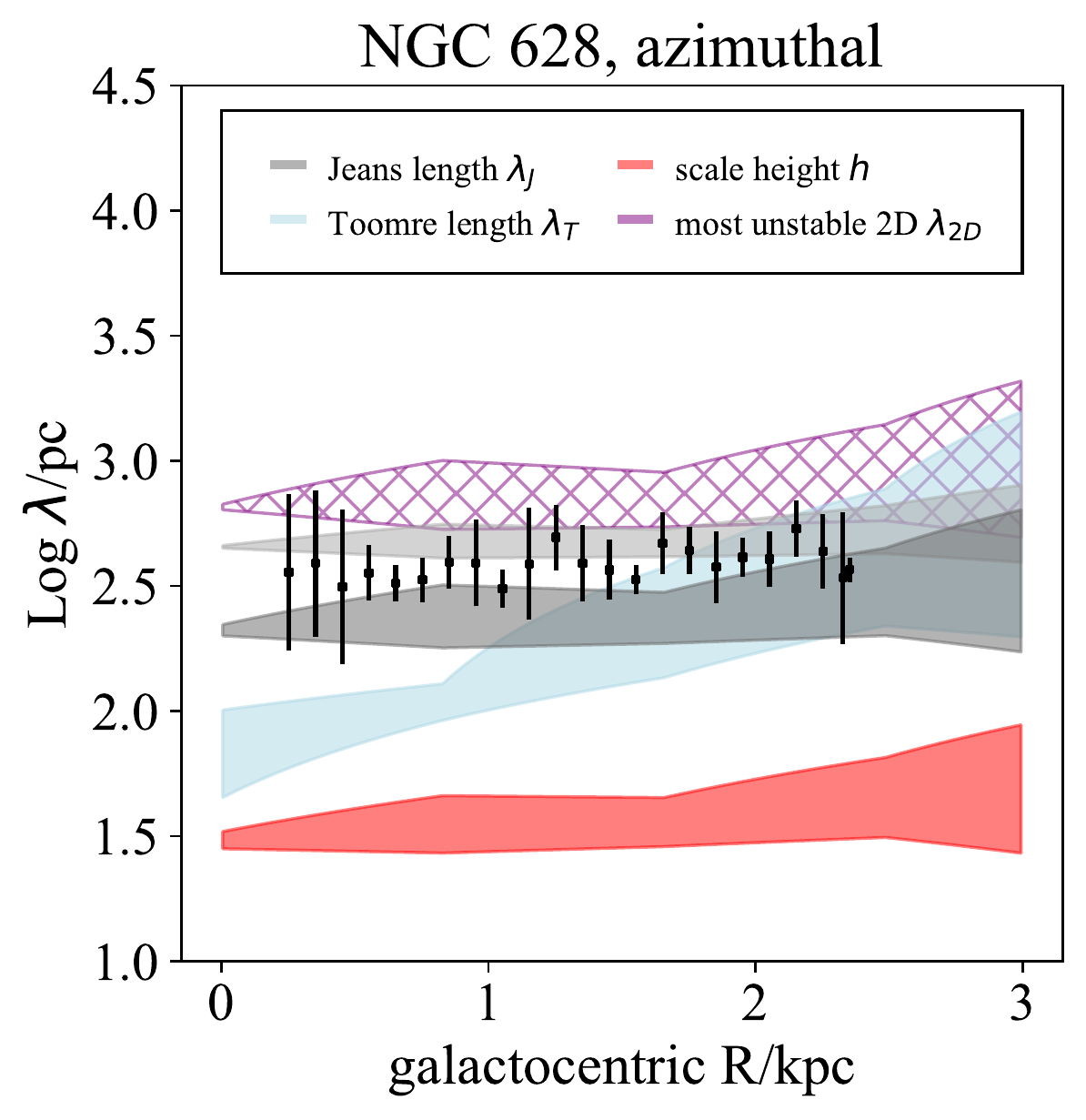}&
\includegraphics[width=.5\linewidth]{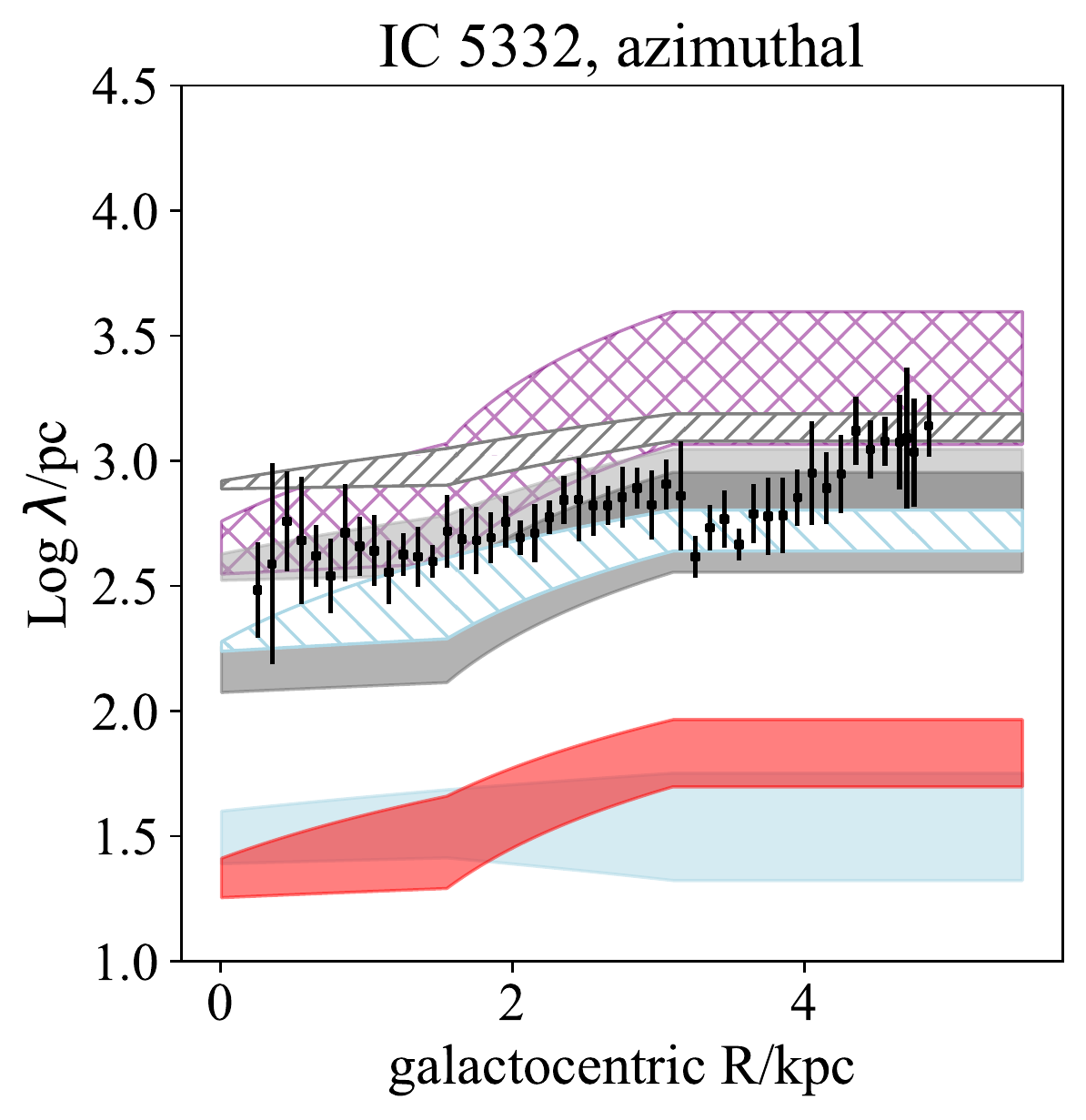}\\
\includegraphics[width=.5\linewidth]{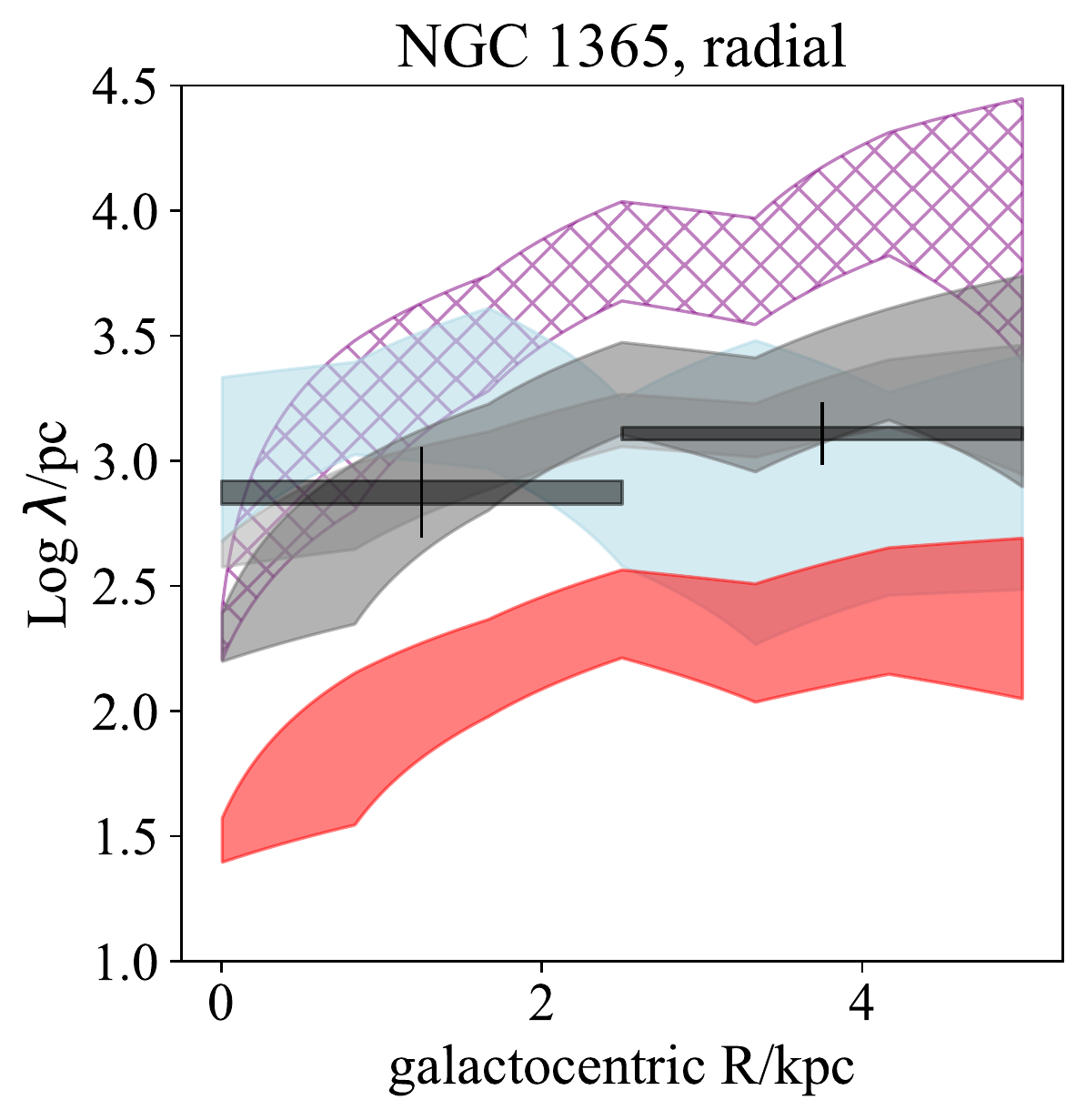}&
\includegraphics[width=.5\linewidth]{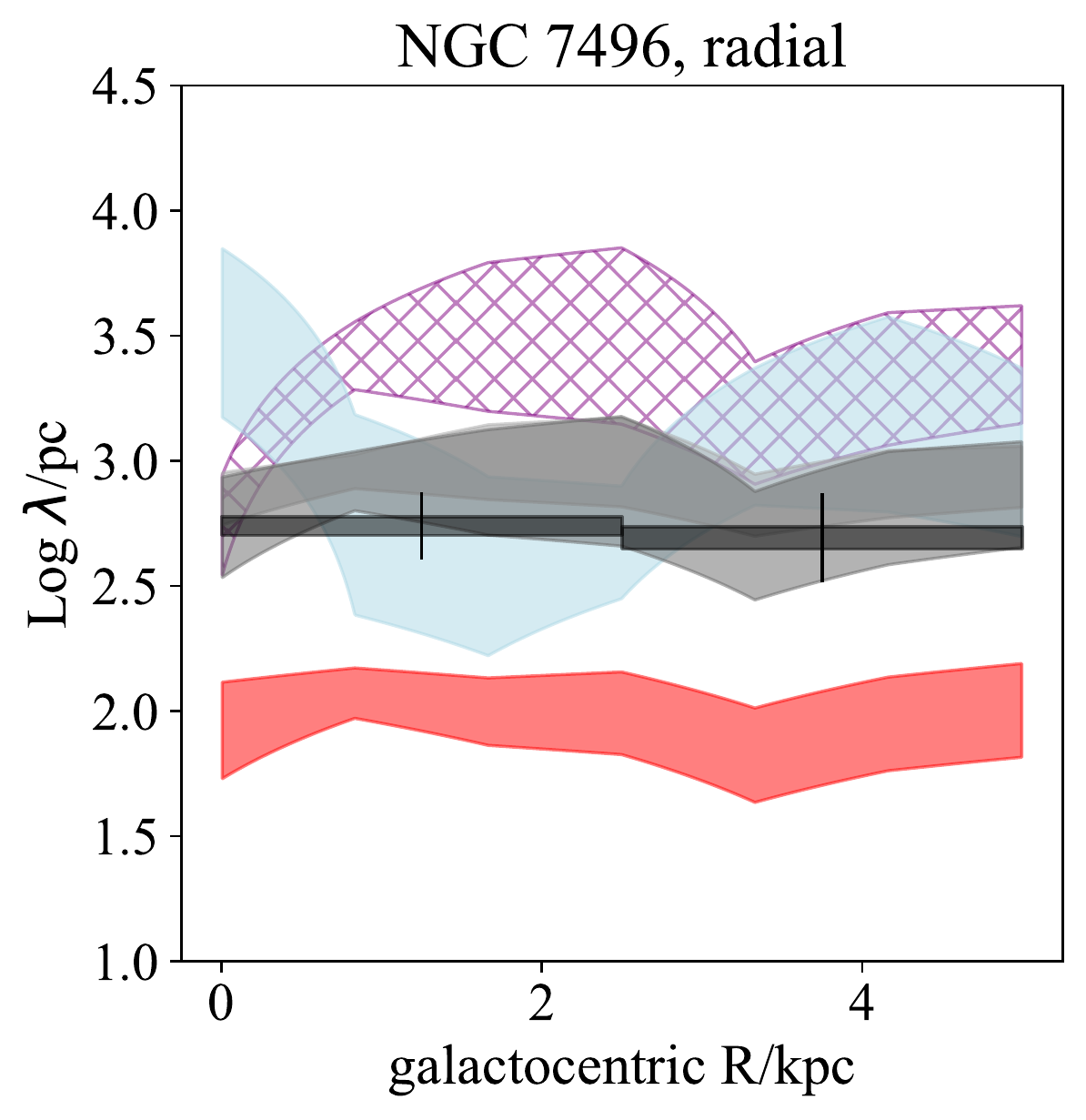}
\end{tabular}
\end{center}
\caption{Comparison between the spacings measured for filaments detected by FilFinder (black) and estimates for the Jeans length, Toomre length, most unstable 2D wavelength and gas scale height (gray, blue, red and purple hatched regions) in (clockwise from top left) NGC~628, IC~5332, NGC~7496 and NGC~1365.  In NGC~628 and IC~5332 the average azimuthal filament spacings in 300-pc wide (projected) radial bins are shown. In NGC~1365 and NGC~7496, the average radial filaments spacings in two radial zones $R<$2.5 kpc and $R>2.5$ kpc are shown.  The dark gray bar depicts the Jeans length estimated (where possible) using a determination of the equilibrium gas scale height show in red (see $\S$ \ref{sec:jeanstoomre}), while the light gray band shows the range of values implied adopting a fixed 100-pc scale height.  The light blue band traces the range in Toomre lengths estimated for each gas disk.  The purple band shows the value estimated for $\lambda_{2D}$ where 2D structure preferentially forms as long as $Q$ does not exceed unity.  Note that, given the estimated $\lambda_T$, these $\lambda_{2D}$ estimates would all coincide with $Q>1$, disfavoring structure on this scale (see eq. [\ref{eq:toomre}]). Additional hatched light gray and light blue bands for IC 5332 show, respectively, the Jeans length (forward hatch) and Toomre length (backward hatch) when an atomic gas component is included, assuming a constant $h$=100~pc.
 }
\label{fig:spacings}
\end{figure*}

\subsubsection{Measuring Characteristic Spacings}
The filamentary features of interest in the MIRI images of our four targets exhibit interesting patterns in their lengths, orientations and separations.  In this section we wish to obtain a preliminary assessment of the basic structure in these images and focus only on filament separations.

The filaments in all targets are elongated in both the radial and azimuthal directions, but visually, the filaments in the two barred galaxies NGC~7496 and NGC~1365 are oriented more in the azimuthal direction than the filaments in NGC~628 and IC~5332.  In practice, this means the separation between neighboring filaments is mostly radial in the first two cases and azimuthal in the latter two cases.  The filaments in NGC~628 and IC~5332 also tend to increase in both separation and length moving outwards in galactocentric radius.  Within the context of disk fragmentation, the relative sizes of radial and angular fragmentation scales can yield insight into the triggers of gravitational instability \citep{meidt22}.  However, the covariance of radial and azimuthal spacings and filament lengths is beyond the scope of the present work and we proceed with the simplest possible spacing measurement for each target, given the typical orientation of the observed filaments.  We thus measure either the azimuthal separation between filament features at a given radius -- in NGC~628 and IC~5332 -- or the radial separations between filaments at a given azimuth -- in NGC~1365 and NGC~7496.  Our approach, described below, misses a direct characterization of the quasi-periodicity visible in the images, but the determination of, e.g., Fourier power spectra might be fruitful in future.

To measure the azimuthal spacings of filaments in the FilFinder masks, we first divide the masks into a series of narrow elliptical annuli projected onto the sky plane at the orientation of the galaxy (see Table \ref{tab:props}).  The width of the annuli is chosen to be one resolution element or 3 pixels in the MIRI images (15 pc at the distance of NGC~628).  Then we find the azimuthal sites of intersection between the filaments in the mask and each ring.  All filament mask objects that intersect the ring with fewer than a threshold of $N_{\rm pix}=5$ pixels are ignored.  These are assumed to be physically uninteresting for the present study. Once the positions of all remaining filaments are recorded, we determine the difference between each pair of neighboring filament-ring intersections in a given ring.  Finally, we measure the mean and standard deviation of angular separations between filament mask objects in each ring.

Given our choice of a relatively fine bin width, a given filament tends to appear in a number of neighboring annuli (depending on the length of the filament).  This makes the number of filament-ring intersections per ring mostly stable from ring to ring, given that, in a given radial zone, the filaments are overall fairly uniform in width and length.  But there are some variations that reflect the finite lengths of filaments and their diverse shapes and widths.  Large variations in filament properties might require a change in the adopted $N_{\rm pix}$.  For now, we let that information contribute to the mean angular distance between filaments and how it varies as a function of galactocentric radius.  We adopt a final radial averaging and measure the mean angular separation in every twenty radial bins, i.e. in 300 pc-wide annuli.  

It is worth noting that this approach does not measure the spacings between spur features adjacent to the spiral arms.  These features contribute typically only one major intersection per annulus, which otherwise predominantly sample filaments at locations away from the spiral arms.  

To measure the radial spacings of the ring-like filaments in the FilFinder masks for NGC~1365 and~NGC7496, a similar procedure is adopted but now the masks are divided into 1 degree wide azimuthal segments rather than annuli.  In each azimuthal segment, the radial intersection between the filaments in the mask and the segment is sought, omitting any intersections consisting of less than $N_{\rm pix}=5$.  Finally, the mean and standard deviation of the nearest neighbor separations is calculated.  To modestly account for the possibility of radial variation in the radial spacings, segments are analyzed in two pieces, $R$$<$2.5 kpc and $R$$>$2.5 kpc. In each of these radial zones we obtain a single representative measure of the filament radial spacing by taking the average of the spacings measured across a limited range of azimuths, where the filamentary features under consideration are most visible (away from the bar ends).

\subsection{Estimating the Turbulent Jeans Length and the Toomre Length}\label{sec:jeanstoomre}
We wish to compare the measured filament spacings with basic predictions for the characteristic fragmentation scale in rotating gas disks.  The well-known Toomre length \citep{toomre}
\begin{equation}
\lambda_T=\frac{4\pi^2G\Sigma}{\kappa^2}\label{eq:toomre}
\end{equation}
is the smallest scale stabilized by rotation in a 2D rotating disk and is typically hundreds of parsecs to a few kpc.  We determine $\lambda_T$ at the locations of the mega-table hexagonal apertures (see $\S$ \ref{sec:almadata}) for each galaxy using the observed gas surface density and an estimate for the radial epicyclic frequency
\begin{equation}
\kappa=\left(4\Omega^2+R\frac{d\Omega^2}{dR}\right)^{1/2}
\end{equation}
derived from an analytical fit to the observed rotation curve \citep{lang20}. Here $\Omega=V_{c}/R$ with $V_c$ the circular velocity (see $\S$\ref{sec:almadata}).  

We also compare with the most unstable 2D wavelength 
\begin{equation}
\lambda_{2D}=\frac{2\sigma^2}{G\Sigma}=2\lambda_{J,2D},\label{eq:2Dmostunstable}
\end{equation}
which marks where the growth of instabilities is most rapid, positioned above the 2D Jeans length $\lambda_{J,2D}$ (where growth is infinitely slow).  In this expression, $\sigma$ is the turbulent plus thermal velocity dispersion in the gas \citep[following][]{chandrasekhar51}.  

Fragmentation occurs at $\lambda_{2D}$ subject to the condition that the Toomre parameter 
\begin{equation}
Q=\frac{\sigma\kappa}{\pi G\Sigma}=\left(\frac{\lambda_{2D}}{2\lambda_T}\right)^{1/2}\leq1. \label{eq:toomrecriterion}%
\end{equation}
When the disk is stable and $Q\gtrsim1$, estimates for $\lambda_{2D}$ exceed $\lambda_T$. Since rotation entirely prevents growth at and above $\lambda_T$, this scenario underlines the 2D stability predicted by $Q\gtrsim1$. 

Accounting for the 3D nature of gas disks, the turbulent Jeans length \citep{jeans}
\begin{equation}
\lambda_J=\frac{\sigma \pi^{1/2}}{(G\rho)^{1/2}}=\lambda_{2D}\frac{f_g^{1/2}}{\pi}\label{eq:jeans}
\end{equation}
is predicted to be the more natural scale for fragmentation at the mid-plane in rotating gas disks \citep{meidt22}.  The relation to $\lambda_{2D}$ in this expression is written in terms of the gas fraction $f_g$=$\rho/(\rho+\rho_b)$ where $\rho_b$ is the background density.  3D instability can occur on this scale even when rotation is predicted to lead to 2D stability with $Q>1$. The presence of rotation slightly lengthens gravitationally unstable fragments \citep{meidt22}, but the Jeans length serves as a good lower bound on the 3D fragmentation scale. 

In the molecular disks of nearby galaxies, the turbulent Jeans length is typically on the order of tens to hundreds of parsecs,  systematically smaller than $\lambda_{2D}$.  In some scenarios, $\lambda_J$ approaches or exceeds $\lambda_T$, i.e. when the Toomre criterion is not satisfied ($Q\gtrsim1$; see eqs. [\ref{eq:2Dmostunstable}] and [\ref{eq:jeans}]), signaling that only 3D instability may be possible.   

With constraints from PHANGS-ALMA for the gas surface density and turbulent velocity dispersion, estimating $\lambda_J$ throughout each galaxy requires a determination of the gas volume density and thus the vertical scale height.  Where possible we do this in two ways: either assuming a uniform 100 pc scale height (i.e. the height of the MW disk, \citealt{heyerdame15}) or solving for the scale height required for hydrostatic equilibrium in the presence of the self-gravity and a background potential, i.e. \begin{equation}
\frac{dP}{dz}=-\rho\int\nabla_z^2\Phi dz
\end{equation}
in terms of the gas pressure $P$=$\rho\sigma^2$ and total gravitational potential $\Phi$ that obeys Poisson's equation $\nabla^2\Phi=4\pi G(\rho+\rho_b)$.  

Assuming that the gas is isothermal and follows a Gaussian vertical distribution $\rho\propto exp(-z^2/(2h^2))$ with scale height $h$ (such that $\Sigma=\rho\sqrt{2\pi}h$) and making the typical assumption that the vertical terms dominate Poisson's equation, this can be written as the approximation
\begin{equation}
\frac{\sigma^2}{h^2}z=\int\left(\sqrt{8\pi}G\frac{\Sigma}{h}+4\pi G \rho_b\right)dz
\end{equation}
yielding the following quadratic equation to solve for $h$,
\begin{equation}
\sqrt{8\pi}G\Sigma h +4\pi G\rho_b h^2-\sigma^2=0.\label{eq:scaleheight}
\end{equation}

For this work, the background is assumed to be dominated by the underlying stellar distribution, omitting the contribution from atomic gas and the dark matter halo (unless otherwise specified).  This may lead to overestimation of the gas scale height and Jeans length determinations, but should be suitable for the preliminary estimates presented in this work.  Following \cite{sun22}, the stellar volume density is estimated from the observed stellar mass surface density together with the galaxy's empirically assigned stellar scale height.  The gas scale heights estimated for our four targets lie typically in the range 30 to 100 pc and exhibit a modest amount of radial variation.  Replacing the nominal velocity dispersion with the conservative $\sigma_{\rm gas}\sim$11 km s$^{-1}$ \citep[following][]{leroy08}, brings this to a typical value of 100 pc on average.  Using either scale height, the mid-plane gas volume density is calculated as $\rho=\Sigma/(\sqrt{2\pi}h)$, assuming a Gaussian vertical distribution.  

\section{Results}\label{sec:results}
Figure \ref{fig:spacings} plots the JWST filament spacing measurements for each of the four galaxies together with estimates for the Toomre length and the Jeans length in the studied regions plus the two other reference lengths $h$ and $\lambda_{2D}$. The vertical error bar on each measurement shows the average standard deviation of spacings in each 300-pc wide radial bin in the upper two panels or within the 1-deg wide azimuthal segments in the lower two panels.  In the case of radial  spacings, the height of the horizontal bars spanning the two radial zones $R<$2.5 kpc and $R>$2.5 kpc shows the standard deviation of the averages.     

The Toomre length, the Jeans length, the scale height $h$ and the most unstable 2D length $\lambda_{2D}$, which are shown as blue, gray, red and hatched purple bands, respectively, are determined where possible using PHANGS-ALMA CO as a gas dynamics tracer. The width of each band highlights the full spread in values at a given radius arising with azimuthal variations in the observed gas properties. 
Two additional hatched bands included in the case of IC 5332 show the Jeans and Toomre lengths when an atomic gas component is included (see $\S$ \ref{sec:almadata}).  The purple band for $\lambda_{2D}$ is shown with hatching as an indication that, given the estimated $\lambda_{T}$, structure would only form at $\lambda_{2D}$ as long as $Q$ does not exceed unity.

For reference, the low side of each gray shaded zone (showing the Jeans length estimates) originates with the higher surface density regions at each radius (e.g. in spiral arms), while the upper side of each band is more representative of lower-density (interarm) regions.  These gray bands would steepen and slightly shift up (by less than 0.2 dex in the case of NGC 7496 and NGC 1365 and by as much as $\sim 0.3-0.5$ dex in NGC 628) with an alternative global gas velocity dispersion $\sigma_{gas}$=11 km s$^{-1}$ \citep{leroy08}, which is a factor of 2-3 higher than our fiducial values on average.  The scale height chosen to calculate $\lambda_J$ makes a comparable change to the estimated Jeans lengths.

In all cases, the filament spacings fall very near the Jeans length and agree less well with the other reference lengths, although the Toomre and Jeans lengths are often similar (as considered later in $\S$ \ref{sec:discussion}).  Consistent with the visual impression from the maps, filaments tend to modestly increase in separation moving outwards in galactocentric radius.  A similar behavior is typical of the Jeans length in these targets, given observed gas properties.  

\begin{figure}[t]
\begin{center}
\vspace*{-.5in}
\hspace*{-1.05cm}
\includegraphics[width=.95\linewidth]{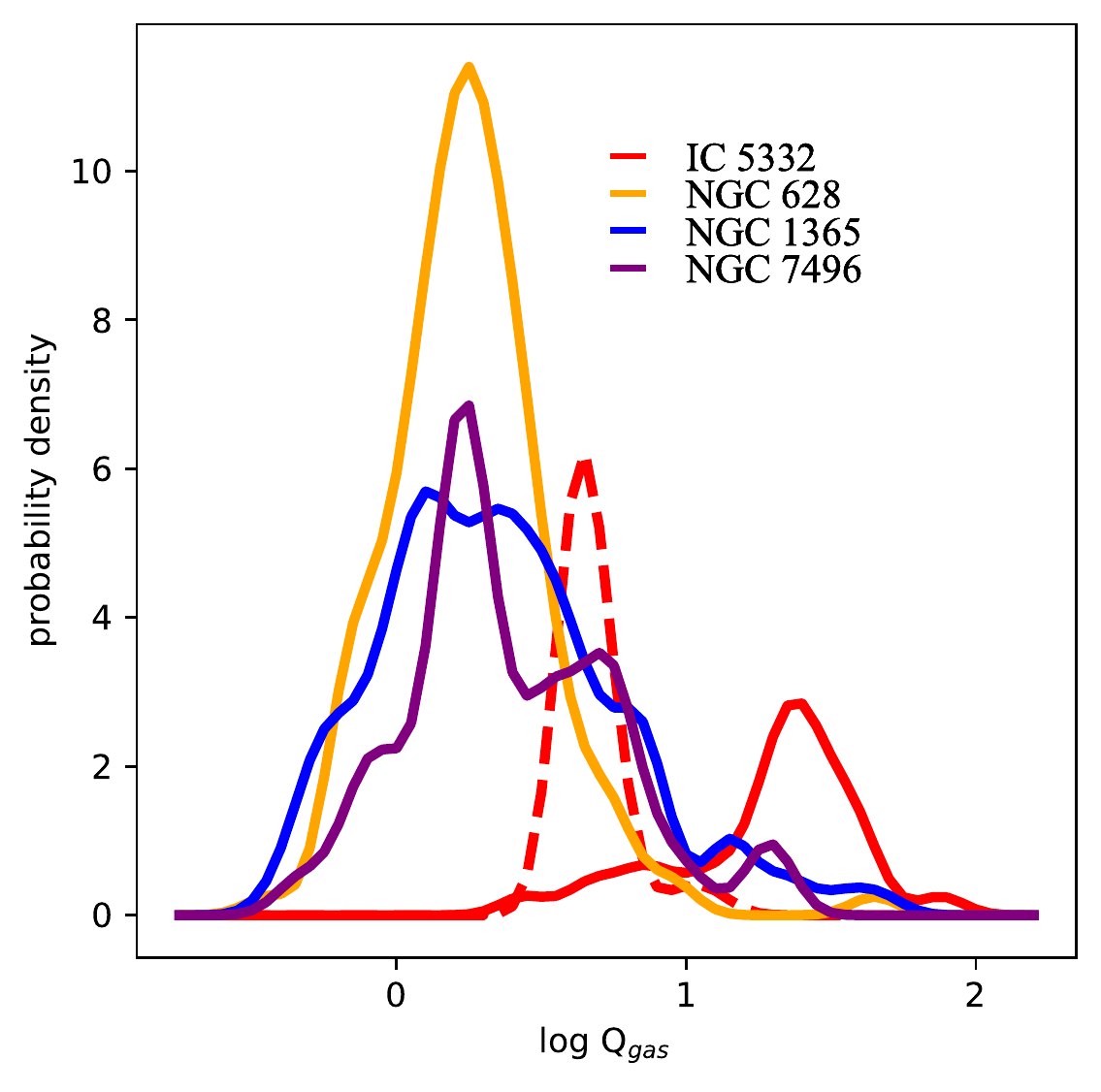}
\end{center}
\caption{KDE histograms of Toomre Q values for the gas in the four target galaxies calculated from observed gas properties across each CO-emitting disk.  The dashed (solid) red line for IC~5332 shows the Q values estimated with (without) an atomic gas component.  \\
 }
\label{fig:Qhist}
\end{figure}

Like the Jeans lengths, the scale height and $\lambda_{2D}$ also tend to modestly increase with $R$ but consistently fall below or above the measured spacings, respectively.  (In non-fully self-gravitating gas disks like those studied in the local universe here, $h$ 
is everywhere systematically smaller than $\lambda_J$, while $\lambda_{2D}>\lambda_J$.) As discussed more below, the structure that forms through the expansion of feedback-driven bubbles (and conceivably a number of other processes) should appear preferentially on or below scales near the disk scale height, which is a natural and imposed length scale for the
system. 
The gas scale height would need to be $\sim$500 pc to match to the observed filament spacings, a highly atypical value for atomic or molecular gas inside 10 kpc \citep[e.g.,][]{heyerdame15}. The 2D length $\lambda_{2D}$ would be more easily brought into agreement with the observed spacings, requiring velocity dispersions 1.8 times lower than the values adopted in this work (which are already conservatively low) or surface densities higher by a factor $\sim$3. However, no structure would be able to form at $\lambda_{2D}$ where $Q$ exceeds unity, as indeed appears to be the case throughout most of the gas disks of all four targets (Figure \ref{fig:Qhist}), as can also be inferred from our finding that typically $\lambda_{2D}\gtrsim\lambda_T$.   

The atomic gas component, which is minimal across the MIRI field of view in all targets in this work except IC~5332, is expected to yield minimal change to the estimated Jeans or Toomre lengths for NGC~628, NGC~1365 and NGC~7496 shown in Figure \ref{fig:spacings}.  However, in other less molecule-rich targets and regions, it can be an important contribution to the gas self-gravity and hence influential for disk stability.  Since we lack precise observational constraints on the atomic gas surface density and velocity dispersion for NGC~1365 and NGC~7496, we do not make a quantitative assessment of stability including the atomic gas here.  Instead we note that, adopting representative values for the atomic gas surface density (10 M$_\odot$ pc$^{-2}$; \citealt{bigielblitz12}) and velocity dispersion ($\sigma$=11 km s$^{-1}$; \citealt{leroy08}), $Q$ is on average 1.5-2 times larger in the atomic gas than in the molecular gas and the atomic gas Jeans length is near $\sim$ 1~kpc, similar to the Jeans lengths estimated here.  This is consistent with the expectation that the molecular and atomic components of the gas are dynamically well-mixed, such that instability takes place in a mix of gas (not only the molecular gas), but preferentially at high density often traced by the molecular phase.

\section{Discussion}\label{sec:discussion}
The filamentary structure on 200-500 pc scales studied in this work presents an interesting inconsistency with basic expectations for structure formation through gravitational instability or star formation feedback.  

\subsection{Filaments as a product of gravitational instability}\label{sec:filamentsinstability}
Traditional theoretical arguments for the development of gravitational instability in rotating disks \citep[e.g.][]{toomre, GLB} predict that rotation slows the growth of structures forming on scales $\lambda>\lambda_{J,2D}$ (with slowest growth on the smallest scales) and entirely suppresses fragmentation above the Toomre length (eq. \ref{eq:toomre}).  
The result is a preference for fragmentation on the intermediate scale $\lambda_{2D}$ coinciding with the Toomre threshold $Q=\sigma\kappa/(\pi G\Sigma)=1$.

In the disks of the present target sample, $\lambda_{2D}$ lies near or above $\lambda_T$, which occurs when $Q\gtrsim1$ (see eq. \ref{eq:toomrecriterion}).  Indeed, the gas disks here (like most gas disks in nearby galaxies) already appear to be generally Toomre stable (see Figure \ref{fig:Qhist} and e.g. \citealt{leroy08} and Williams et al., in prep.).  This makes 2D fragmentation near $\lambda_{2D}$ or any scale above $\lambda_T$ highly unlikely.  

In this case, the structure on scales below $\lambda_{2D}$ but above or near to $\lambda_T$ found in this work would suggest that fragmentation may be occurring in a mode in which factors are active that reduce the influence of rotation in favor of gravity.  For example, in magneto-Jeans instability (MJI) magnetic forces can act to counter the stabilizing influence of rotation \citep{elm87, ko01}.  However, this is most effective in diffuse ISM phases where the Alfv\'{e}n speed is high compared to turbulent plus thermal velocity dispersion.  In the cold neutral gas with characteristic high densities and equilibrium pressures \citep{sun20}, the Alfv\'{e}n speed is low and magnetic forces will mostly act as a source of stability \citep{ko01}, disfavoring structures near $\lambda_{\rm J,2D}$ and lowering the stability threshold.  Given the $Q$ values estimated for these targets (Figure \ref{fig:Qhist}) magnetic forces would even more strongly disfavor 2D fragmentation near $\lambda_{2D}$.  

Turbulent dissipation \citep{elm11} has been proposed as an avenue for shifting (and possibly removing) the Toomre condition that can also lead to growth below $\lambda_{2D}$.  Taking into account thickness, the stability threshold in 2D rotating gas disks with dissipation shifts to $Q\sim2$, allowing for weak instability above $Q$=1.  The instability in this dissipation-assisted regime is characterized by a fastest growing length that shifts to smaller scales as $Q$ increases, though with growth rates that are typically an order of magnitude lower than when $Q<1$.

Another way that gravity has been shown to gain a strong hold over rotation is in a fully 3D scenario, including the additional degrees of freedom associated with the third (vertical) direction  \citep{meidt22}.  Accounting for the perturbed vertical pressure and gravitational forces absent from 2D treatments, 3D fragmentation can be triggered even where 2D fragmentation is suppressed, subject to a modified stability threshold 
 \begin{equation}
Q_M=\frac{\kappa^2}{4\pi G\rho}=\sqrt{\frac{\pi}{2}}\left(\frac{\lambda_J}{\lambda_T}\right)f_g^{1/2}\equiv 1,\label{eq:QM}
\end{equation}
or roughly $Q=2$, again writing $f_g$=$\rho/(\rho+\rho_b)$.  The 3D structures that form in this scenario are able to grow rapidly, as fast as Toomre instabilities, down to scales $\lambda_J\lesssim\lambda_{2D}$ 
(\citealt{meidt22}; taking into account the smaller scale height in the presence of a background potential with density $\rho_b$ as in eq. [\ref{eq:jeans}]).  Even where the disk is stabilized in 2D by rotation at and above $\lambda_T$, 3D instability can occur near $\lambda_J$ up to the threshold $Q_M=1$ where $\lambda_J\approx\lambda_T$ (see eq. [\ref{eq:QM}]). 

From this perspective, the observed gas structures on hundreds of pc scales suggests that disk fragmentation can indeed occur in a mode in which shear plays little role compared to pressure and gravity, as predicted \citep{elm87,ko01,elm11,meidt22}.  Rotation could nevertheless be important for regulating the formation of structure on larger scales. 

Previously, Jeans scale fragmentation had been proposed for describing the spacing of clumps in low shear spiral arm filaments \citep{elm18}.  For the filaments that form a disk-wide network and originate with disk fragmentation, the next levels of fragmentation might be scale-dependent and ending at the thermal Jeans length, according to dissipation and the scale-dependence of turbulence \citep[as observed in MW filaments; e.g.][]{mattern18} and also sensitive to stability in the new filamentary geometry \citep[e.g.][]{nagasawa87,freundlich14}. 
The internal kinematics of these structures should be studied in order to probe how subsequent fragmentation can proceed, eventually leading to star formation.  In this context we note that, although the formation of small-scale filamentary structures through disk fragmentation may not be sensitive to shear, they are not necessarily insensitive to galactic orbital motions: the growth rates of 3D fragments still depend on shear \citep{meidt22} and the turbulent velocity dispersion responsible for the gas pressure that counters self-gravity partially reflects (orbital) motion in the galactic potential in equilibrium \citep{meidt18,sun20}.  Structures formed through 3D disk fragmentation may thus not be kinematically decoupled from their host galaxy.  

\subsection{Filaments in relation to feedback}
As alluded to already in $\S$ \ref{sec:results}, the filamentary structure evident in the MIRI images is spaced with a characteristic scale that is 1-1.5 orders of magnitude larger than the disk scale height.  Often spatially coincident with these filaments, there is an abundance of rich structure evident on smaller scales, with shell-like and clumpy morphologies \citep{WATKINS_PHANGSJWST}, that can be linked to directly to star formation \citep{WATKINS_PHANGSJWST,BARNES_PHANGSJWST}.  Indeed, the MIRI-traced filamentary web harbors 75--80\% of HII regions and 60\% of star clusters younger than 5 Myr in NGC~628 \citep{THILKER_PHANGSJWST}.  

The planned detailed studies of star formation feedback leveraging the full multi-wavelength PHANGS data set will be key for understanding how feedback-driven structures form and evolve and interact with structures that are primarily dynamical in origin.  
We propose that the scale height is a useful reference for distinguishing gas structures that are dynamical in origin from those that are feedback-driven.  Theory and numerical simulations predict that many bubbles driven by SNe feedback can stall at or below $h$ \citep[whether in an energy- or momentum-driven regime;][]{maclow88, kim17,fielding18,orr22} and even the bubbles that do break out of the plane may expand only slightly beyond $h$ \citep[considering that wind momentum goes to driving a vertical outflow rather than to in-plane expansion; see][]{orr22}.  
Consistent with this expectation, the catalog of bubble candidates in NGC~628 morphologically identified in the MIRI images and visually matched with a stellar counterpart \citep{WATKINS_PHANGSJWST} have average radii 20-30 pc, just at the scale height estimated here (see Figure \ref{fig:spacings}), although they can reach as large as 550 pc.  In total, 60\% of the bubbles identified in NGC~628 have radii below 1$h$ and $\sim$ 90\% are within 2$h$.  

With some exceptions \citep[like the `Phantom Void';][]{BARNES_PHANGSJWST}, many of the feedback-driven shells traced by MIRI appear to be stalling at or near the scale height, supporting the idea that structures present on larger scales (with separations $>$ 100 pc) are mostly dynamical in origin.  

In this light, we speculate that the structure originating with dynamical factors like disk fragmentation and shear acts as a scaffold that organizes star formation and feedback, which then shapes the large scale structure and influences the process of disk fragmentation. 
Feedback acts
not only to break up fragments (e.g. most clearly in NGC~7946 and NGC~628; see also \citealt{WATKINS_PHANGSJWST}; \citealt{BARNES_PHANGSJWST}), but serves as a coordinated source of perturbations that stimulate further fragmentation.  From the visual appearance of the four targets studied here, and taking the present distribution of star formation as representative of the time-averaged activity in the recent past, star formation seems to prompt fragmentation with both radial and azimuthal components (e.g. NGC~628 and NGC~7496).  In contrast, ring-like fragmentation may be sustainable in the absence of star formation or other sources of azimuthal perturbations (e.g. within the bar in NGC~1365).  Studies in a greater number of targets will be important for recognizing how certain types of perturbations (local feedback vs. underlying large-scale patterns) act to trigger disk fragmentation with different radial and azimuthal properties.

\subsection{A characteristic scale?}
Spatial power spectra \citep[e.g.][]{elmscalo04,grisdale17,koch20} can be expected to provide an indispensable view of the rich multi-scale web of gas structures created by gravitational instability and feedback visible in the MIRI images.  However, the large scale structure of the galaxy and the resolution of the observations can masquerade as characteristic features in power spectra \citep{koch20}.  The high spatial resolution of the MIRI data are crucial for testing whether there are genuine features in the power spectra associated with the scale height as previously proposed \citep{lazarian,elm01,dutta09,combes12}.  Characteristic Jeans-scale filamentary structure may also leave an imprint on the shape of a power-law power spectrum, but given that the structure follows a preferred radial-azimuthal organization, this signature is perhaps best recovered with a non-Cartesian (non-sky-plane) coordinate system.  Spatial variation in the Jeans length and scale height also calls for radial and/or azimuthal segmentation of the input maps \citep[e.g.][]{elm01}. 

\section{Summary}
High resolution JWST/MIRI images of  PAHs coupled to the cold dense gas in galaxies provide a record of the dynamical processes that shape the way gas is structured and organized across scales.  This study, combined with the work of \cite{WATKINS_PHANGSJWST} and \cite{BARNES_PHANGSJWST}, suggests that a combination of feedback from star formation and large-scale disk dynamics are important for organizing the gas into a web of bubbles, filaments and voids present on tens to hundreds of pc scales in the MIRI images.

In an initial sample of four galaxies, we focus on the subset of filamentary gas structures that most closely resemble the `giant filaments' observed in the MW \citep[see][and references therein]{hacar22}.  These structures are remarkable for their quasi-regularity (in length, width, orientation and spacing) and their ubiquity, appearing not only over large regions in the individual targets, but also in galaxies with diverse morphologies, star formation rates, gas contents and rotational properties.

We present a pilot study of the characteristic spacings between the observed filamentary gas structures and find that they are compatible with the local turbulent Jeans length constrained by observations.  In three of the four targets where the ISM is molecule-dominated across the MIRI field of view, the observed gas properties are traced in the molecular phase, while in the molecule-poor IC~5332, the gas properties are constrained from the combination of molecular and atomic gas.  

The observed filamentary structure is found on scales well above the disk scale height, which might limit the sizes of many feedback-driven structures \citep{maclow88,orr22}, and is present even where the gas is Toomre stable \citep[see also][]{henshaw20}.   

This may be a sign that the observed structure originates as the product of 3D disk fragmentation predicted to occur on the Jeans length at the mid-plane in rotating gas disks, i.e. regulated by the balance between turbulent pressure and self-gravity, without strong sensitivity to shear (\citealt{meidt22}; and see e.g. \citealt{elm87, ko06,elm11}).  In this context, the good agreement between the observed filament spacings and the Jeans length suggests the interesting possibility of inferring the gas density and scale height from observed gas structure.  

The prominence of structure on the Jeans scale highlighted in this work also points to this as the fundamental outer-most scale of the process of hierarchical collapse that leads to star formation.  Indeed, the Jeans length is similar to the characteristic separation length of regions measured in the statistical reconstruction of the cloud lifecycle in these four targets and others \citep{kruijssen19,kim22}.  

Star formation also has its own observable impact on gas structure \citep{WATKINS_PHANGSJWST,BARNES_PHANGSJWST}.  Future work (i.e. leveraging the PHANGS combination of HST, JWST, MUSE and ALMA data) is necessary to study how feedback coordinates with galactic dynamical processes to produce multi-scale gas structure.

\section*{Acknowledgments}
We thank the anonymous referee for a timely and constructive report that improved the quality of this paper.

This work is based on observations made with the NASA/ESA/CSA JWST. The data were obtained from the Mikulski Archive for Space Telescopes at the Space Telescope Science Institute, which is operated by the Association of Universities for Research in Astronomy, Inc., under NASA contract NAS 5-03127. The observations are associated with JWST program 2107 and be accessed via \dataset[10.17909/9bdf-jn24]{http://dx.doi.org/10.17909/9bdf-jn24}. 

This paper makes use of the following ALMA data, which have been processed as part of the PHANGS-ALMA CO(2-1) survey: \\
\noindent ADS/JAO.ALMA\#2012.1.00650.S, \linebreak 
ADS/JAO.ALMA\#2013.1.01161.S, \linebreak 
ADS/JAO.ALMA\#2015.1.00956.S, \linebreak 
ADS/JAO.ALMA\#2017.1.00886.L, \linebreak 
ADS/JAO.ALMA\#2018.1.01651.S. \linebreak 
ALMA is a partnership of ESO (representing its member states), NSF (USA), and NINS (Japan), together with NRC (Canada), NSC and ASIAA (Taiwan), and KASI (Republic of Korea), in cooperation with the Republic of Chile. The Joint ALMA Observatory is operated by ESO, AUI/NRAO, and NAOJ. The National Radio Astronomy Observatory is a facility of the National Science Foundation operated under cooperative agreement by Associated Universities, Inc.

RSK, EJW and SCOG acknowledge funding from the European Research Council via the ERC Synergy Grant ``ECOGAL'' (project ID 855130), from the Deutsche Forschungsgemeinschaft (DFG) via the Collaborative Research Center ``The Milky Way System''  (SFB 881 -- funding ID 138713538 -- subprojects A1, B1, B2, B8 and P2) and from the Heidelberg Cluster of Excellence (EXC 2181 - 390900948) ``STRUCTURES'', funded by the German Excellence Strategy. RSK also thanks the German Ministry for Economic Affairs and Climate Action for funding in the project ``MAINN'' (funding ID 50OO2206). G.A.B. acknowledges support from ANID Basal projects FB210003.  
MC gratefully acknowledges funding from the DFG through an Emmy Noether Research Group (grant number CH2137/1-1). COOL Research DAO is a Decentralized Autonomous Organization supporting research in astrophysics aimed at uncovering our cosmic origins. JMDK gratefully acknowledges funding from the European Research Council (ERC) under the European Union's Horizon 2020 research and innovation programme via the ERC Starting Grant MUSTANG (grant agreement number 714907).  JPe acknowledges support by the DAOISM grant ANR-21-CE31-0010 and by the Programme National ``Physique et Chimie du Milieu Interstellaire'' (PCMI) of CNRS/INSU with INC/INP, co-funded by CEA and CNES.  MB acknowledges support from FONDECYT regular grant 1211000 and by the ANID BASAL project FB210003. FB would like to acknowledge funding from the European Research Council (ERC) under the European Union’s Horizon 2020 research and innovation programme (grant agreement No.726384/Empire)
ER acknowledges the support of the Natural Sciences and Engineering Research Council of Canada (NSERC), funding reference number RGPIN-2022-03499.
JS acknowledges the support of NSERC through a Canadian Institute for Theoretical Astrophysics (CITA) National Fellowship.
KG is supported by the Australian Research Council through the Discovery Early Career Researcher Award (DECRA) Fellowship DE220100766 funded by the Australian Government. 
KG is supported by the Australian Research Council Centre of Excellence for All Sky Astrophysics in 3 Dimensions (ASTRO~3D), through project number CE170100013. 

\typeout{}

\suppressAffiliationsfalse
\allauthors
\end{document}